# Information retrieval in single cell chromatin analysis using TF-IDF transformation methods

Mehrdad Zandigohar
Department of Biomedical Engineering
University of Illinois Chicago
Chicago, Illinois
mzandi2@uic.edu

Yang Dai*
Department of Biomedical Engineering
University of Illinois Chicago
Chicago, Illinois
yangdai@uic.edu

*Abstract*—Single-cell sequencing assay for transposase-accessible chromatin (scATAC-seq) assesses genome-wide chromatin accessibility in thousands of cells to reveal regulatory landscapes in high resolutions. However, the analysis presents challenges due to the high dimensionality and sparsity of the data. Several methods have been developed, including transformation techniques of term-frequency inverse-document frequency (TF-IDF), dimension reduction methods such as singular value decomposition (SVD), factor analysis, and autoencoders. Yet, a comprehensive study on the mentioned methods has not been fully performed. It is not clear what is the best practice when analyzing scATAC-seq data. We compared several scenarios for transformation and dimension reduction as well as the SVD-based feature analysis to investigate potential enhancements in scATAC-seq information retrieval. Additionally, we investigate if autoencoders benefit from the TF-IDF transformation. Our results reveal that the TF-IDF transformation generally leads to improved clustering and biologically relevant feature extraction.

## I. INTRODUCTION

Single-cell sequencing assay for transposase accessible chromatin (scATAC-seq) measures DNA accessibility within the whole genome to reveal regions of accessible chromatin at the cellular level [1] and facilitates the identification of cis-regulatory elements (e.g., enhancers and promoters) in heterogeneous cells. After initial bioinformatics analysis, accessible regions are identified as peaks. If a sequence read from a cell aligned in a peak region, then the corresponding region in that cell is considered accessible and labeled as 1, otherwise labeled as 0. Through this process, the scATAC-seq data from a study is summarized as a sparse binary matrix with columns representing cells and rows expressing the accessible regions (peaks). This matrix is used for downstream analysis for clustering cells with similar chromatin-accessible profiles. However, the high dimensionality and sparsity of the matrix present challenges in identifying cell states based on the scATAC-seq data alone [2]. Thus, some researchers proposed to integrate scATAC-seq data with single-cell transcriptomic sequencing (scRNA-seq) and then transfer the labels from the annotated scRNA cells to the scATAC cells using methods such as anchor-based integration for label transfer, including Signac and MAESTRO for either matched cells or cells profiled from the same biological condition [3-5]. However, the cell states identified by this type of approach are dominated by information from the scRNA-seq data. Thus, it may not be able to capture epigenetic heterogeneity in the cells. Therefore, there is a need to develop effective methods to retrieve essential information (peak features) to capture cellular heterogeneity at chromatin from scATAC-seq data.

Term-frequency inverse-document frequency (TF-IDF) transformation as a normalization method has been utilized in some pipelines for scATAC-seq data analysis to partially address the sparsity issue [3, 6, 7]. TF-IDF transformation methods were first proposed in natural language processing to highlight important features, known as "terms" and demote insignificant ones in a corpus of "documents" [8]. In the scATAC-seq data analysis, a TF-IDF transformation is first applied on the binary peak matrix, followed by a dimension reduction technique, typically singular value decomposition (SVD), prior to cell clustering and downstream analysis. These procedures are generally named latent semantic indexing (LSI), also referred to as latent semantic analysis.

While SVD is effective, it may be computationally expensive for a huge data matrix. Thus, truncated SVD approaches in which the number of components in SVD is limited for downstream clustering analysis are generally utilized as a standard solution. For example, the R package *irlba* has been developed based on a fast and memory-efficient algorithm for SVD truncation [9]. In addition to SVD, other dimension reduction methods, such as matrix factorization (or factor analysis) and autoencoders, have been proposed to generate a latent representation of a binary peak matrix without latent semantic indexing. In a recent study [10], researchers performed a comparative analysis on scATAC-seq cell state detection based on various autoencoders (i.e., autoencoder, sparse autoencoder, variational autoencoder, and stacked autoencoder) and two matrix factorization methods (i.e., non-negative matrix factorization and alternating non-negative least squares matrix factorization). However, their comparison did not include SVD-based methods and did not use any LSI transformation. Thus, to users, it is not clear what is the best procedure to use in order to identify the cell states. Additionally, the current methods incorporating LSI do not facilitate the analysis of feature importance, i.e., important genomic regions, which are essential to elucidate the distinct cis-regulatory elements underlying the heterogeneity of cells.

* Corresponding author
Support of NIDDK DiaComp (DK076169) is acknowledged.

Despite of potential of LSI in tackling scATAC-seq analysis challenges, to the authors' best knowledge, there has not been a comprehensive evaluation of various combinations of the available TF-IDF methods and the dimension reduction techniques on the efficacy and accuracy in defining cell states. Thus, we conducted an investigation focusing on the following five aspects:

(1) how effective the TF-IDF methods (5 methods) (see Table I) are in cell state identification and if a specific transformation technique outperforms the rest;

(2) how the choice of different dimension reduction techniques affects the cell state identification by comparing the SVD-based method, matrix factorization models, and autoencoders;

(3) how the choice of a TF-IDF transformation affects each of these dimension reduction methods;

(4) if these methods are scalable and feasible on large datasets; and lastly

(5) if the LSI method retains important features (genomic regions) in each cell type using Gene Ontology (GO) enrichment and transcription factor motif enrichment analyses.

Using two biologically well-annotated scATAC-seq datasets, our analysis shows that TF-IDF methods are promising in scATAC-seq data analysis; most effective on SVD-based and matrix factorization methods. Autoencoders perform better than other linear methods, but the impact of TF-IDF transformation on them is inconclusive. On the other hand, autoencoders are relatively memory-demanding and time-consuming. They require a careful choice of hyperparameter for model tuning, which could be a downside to non-experts in deep learning. Furthermore, we analyzed the specific SVD components obtained from the TF-IDF transformed data. We found that the LSI can extract biologically relevant genomic regions, demonstrating the effectiveness of TF-IDF for downstream analysis.

## II. METHODS

Two datasets from human and mouse studies were used for the analysis in this paper. Five TF-IDF methods (Table I) in eight scenarios and seven dimension reduction methods: SVD, NMF, lsNMF, general autoencoder, sparse autoencoder, variational autoencoder, and stacked autoencoder were compared.

### A. Datasets

Two scATAC-seq datasets were employed in our evaluation. The cells in both datasets have already been annotated and used as the gold standards for our performance evaluation. The first dataset [11] of 2088 cells generated from mouse forebrain cells (GSE100033) was the primary dataset for our comparative analysis through this paper. It contains 8 cell types: excitatory neuron cells (EX1, EX2, and EX3), inhibitory neuron cells (IN1 and IN2), astrocytes (AC), oligodendrocyte (OC), and microglia (MG). The second dataset was used to investigate scalability and consists of 5 times more cells. The dataset was obtained from the 10X multiome human peripheral blood mononuclear cells (PBMC) from a healthy donor with granulocytes removed through cell sorting (10k cells). This dataset is available on the 10X Genomics website. After excluding the rare cell types (<100 cells), the dataset consists of 16 different cell types: CD4 Naïve, CD4 TCM, CD4 TEM, CD8 Naïve, CD8 TEM, MAIT, NK, gDT, pDC, Treg, B Naïve, B intermediate, B memory, CD14 Mono, CD16 Mono, and cDC2.

The binary peak matrix of the first dataset has been processed in [10]. Specifically, peaks were removed if they contain ≥2 reads in less than 6% of the cells. The second dataset was filtered based on the Signac manual [5] and binarized as 1 for reads greater or equal to 1 and 0 for the rest.

### B. TF-IDF methods

In the TF-IDF methods, the genomic regions are terms, and the cells are documents. The term frequency is defined as the number of peaks (binary) of a region in a given cell over the total number of peaks in the cell. The inverse document frequency is defined as the inverse of the number of peaks in a given peak region across all the cells. More precisely, given a peak $i$ in a cell $j$, and a total population of $N$ cells, we define TF and IDF as below:

$$TF_{i,j} = \frac{f_{i,j}}{\sum_i f_{i,j}} \quad (1)$$

$$IDF_i = \frac{N}{\sum_j f_{i,j}} \quad (2)$$

where $f_{i,j}$ is the binary value (is there is at least 1 sequence read corresponding to the peak $i$ in cell $j$ or not). We investigated 5 different transformation methods as proposed in Table I.

TABLE I. TF-IDF TECHNIQUES IN THE STUDY

| Method | Equation | Ref. |
|---|---|---|
| 1 | IDF | [6] |
| 2 | $log(TF \times IDF)$ | [3] |
| 3 | $log(TF) \times log(IDF)$ | [12] |
| 4 | $TF \times log(IDF)$ | [7] |
| 5* | $log(TF + 1) \times \left(1 + \frac{\sum p_{ij} \log p_{ij}}{\log N}\right)$, $p_{ij} = TF_{ij} \times IDF_i / N$ | [13] |

\* Entropy-based method

### C. Truncated SVD

Following the notations in [9], for a given matrix A, the SVD consists of three matrices:

$$A_{i,j} = U_{i,n_u} D_{n_u,n_v} V_{n_v,j} \quad (3)$$

where $A$ is the binary matrix from a scATAC-seq dataset, $U$ is the matrix of left singular vectors called feature loading matrix, $D$ is the diagonal matrix of singular values, and $V$ is the matrix of right singular vectors named as cell loadings. The important difference between SVD and truncated SVD is, in the latter, the 3 matrices are approximated using $n_u$ and $n_v$ lower dimensions instead of the true dimension size $n_i, n_j$ where $max(n_u, n_v) \leq min(n_i, n_j)$. For our analysis, we used the default number of components as $n_u = n_v = 50$, which is the number of actual LSI components used in the clustering analysis.

*D. Feature extraction*

The feature/cell loadings extracted from the truncated SVD were used for feature analysis. The magnitude of loadings was first scaled based on each LSI component using min-max normalization for easier interpretability inspired by the standard process done in factor analysis for feature analysis.

$$\bar{u}_{ij} = \frac{u_{ij} - \min(u_{.j})}{\max(u_{.j}) - \min(u_{.j})} \quad (4)$$

$$\bar{v}_{ij} = \frac{v_{ij} - \min(v_{i.})}{\max(v_{i.}) - \min(v_{i.})} \quad (5)$$

where $\bar{u}_j$ is the scaled feature loading vector for cell $j$ and $\bar{v}_i$ is the scaled cell loading vector for locus $i$. Then, the top features in each LSI component were selected as important regions contributing to the specific component for downstream analysis.

*E. GO enrichment and motif analysis*

GO enrichment analysis was performed on the top features (loci) of each LSI component to investigate the relevance of each LSI component based on GREAT [14]. The genomic ranges of the top loci for a specific LSI component were drawn out to make bed files. The setting for the association of genomic regions with genes was based on the basal plus extension model, and was set to 1.0 kb upstream and 0.1 kb downstream of the transcription start site (TSS) for proximal regions, plus up to 200 kb for distal regions. Motif analysis was performed based on the Signac instruction. Full genome sequences of the UCSC Mus musculus version mm10 were used. The position frequency matrices (PWM) used for motif analysis were obtained from the JASPAR database (ver. 2020) for the core collection group and vertebrates taxonomy group. The top 300 regions for a given LSI component from the scaled loci loading vectors were used to find enriched motifs as well as the function of cis-regulatory regions in GREAT. MSigDB pathway analysis and GO cellular component provided in GREAT were used for further analysis.

*F. Matrix factorization*

Two matrix factorization methods: (1) non-negative matrix factorization (NMF) and (2) alternating non-negative least squares matrix factorization (lsNMF), were implemented in Python using [10]. These matrix factorization methods decompose the binary matrix into two matrices:

$$A_{i,j} = W_{i,p} H_{p,j} \quad (6)$$

where all matrices are non-negative and $p < max(n_i, n_j)$ denotes the reduced number of components in dimension reduction, and $W$ matrix was used for further analysis. The major difference between the two mentioned methods is, in lsNMF, one of $W$ or $H$ can be non-convex. The other can be found using the least square algorithm.

*G. Autoencoder*

Four autoencoders were examined by modifying the Pytorch script in [10]: (1) simple autoencoder (AE), (2) sparse autoencoder (sparseAE), (3) variational autoencoder (VAE), and (4) stacked autoencoder (stackedAE). Specifically, in sparse and variational autoencoders, the encoders activation functions are all rectified linear units (ReLU), and decoders use Sigmoid function, and Adam's for the optimizer. For the stacked autoencoder, ReLU was used for both encoder and decoder, and the stochastic gradient descent method (SGD) was used for optimization.

The total number of latent features is set to 20 and epochs to be 20,000. Other important hyperparameters are inherited from [10], and readers are referred to the corresponding paper for details. The latent features (data with reduced dimension) from the autoencoders were utilized for clustering.

*H. Clustering*

To find the best TF-IDF transformation on the SVD-based reduced data, the smart local moving algorithm (SLM) for modularity optimization was used on the k-nearest neighbor graph to determine clusters following in the first section of the analysis described in Signac. The clustering method is dependent on a user-defined resolution to obtain the correct number of clusters, which is expected to be equal to the number of cell groups. After selecting the best transformation method, the K-means algorithm was used to group the cells. The number of centroids for the algorithm was selected based on the total number of true unique cell labels, which is 8 for the mouse data and 16 for PBMC.

*I. Performance evaluation*

To evaluate the performance based on the predicted cell cluster labels and the true labels, three metrics were computed: adjusted rand index (ARI), normalized mutual information (NMI), and F1-score. Jaccard similarity was also computed, and heatmaps of the similarity were plotted.

*1) Adjusted Rand Index:* To compare the clustering between the true labels $U$ from gold-standard data, and predicted labels $V$ (by clustering), ARI is computed as below [15]:

$$ARI(U,V) = \frac{2(N_{00}N_{11} - N_{01}N_{10})}{(N_{00}+N_{01})(N_{01}+N_{11}) + (N_{00}+N_{10})(N_{10}+N_{11})} \quad (7)$$

where the number of pairs in different clusters for both sets $U$ and $V$ ($N_{00}$) and the number of pairs in the same clusters for both sets ($N_{11}$) show concordance between $U$ and $V$. On the other hand, the two other variables $N_{01}$ which is the number of pairs within the same cluster in U but not the same in V, and $N_{10}$ the number of pairs within the same cluster in V but not the same in U, measure discordance between the two sets. The upper bound for ARI is 1, which indicates a perfect match between the two sets.

*2) Normalized Mutual Information:* NMI is a measure to estimate clustering quality. It accounts for the quantity of information you can extract from one set, given the other set. The normalization makes it comparable to other NMIs, hence providing a global metric for evaluation. It can be formulated as below [16]:

$$NMI(U,V) = \frac{I(U,V)}{\sqrt{H(U)H(V)}} \quad (8)$$

where $I(U,V)$ is the mutual entropy between the sets $U$ and $V$, and H is the Shannon entropy.

*3) F1 Score:* It aggregates the model precision and recall as one unified metric:

$$F1 = \frac{2 \times (\text{precision} \cdot \text{recall})}{\text{precision} + \text{recall}} \quad (9)$$

It can range from 0 to 1, from the model misclassifying all the cells and 1 correctly classifying all the cells. Where precision $= \frac{TP}{TP+FP}$ and recall $= \frac{TP}{TP+FN}$.

*4) Jaccard Similarity:* A Jaccard similarity coefficient is a measure of similarity between two sets as below:

$$J(U,V) = \frac{|U \cap V|}{|U \cup V|} \quad (10)$$

A Jaccard similarity index of 1 means the two sets are the same, and 0 means there is no similarity.

III. RESULTS AND DISCUSSION

*A. TF-IDF methods enhance cell state detection in SVD-based approaches*

Eight different scenarios of TF-IDF transformation on the mouse forebrain data were performed to investigate the effect of the methods for cell-state detection, as well as the correlation of the first three LSI components with sequencing depth. Fig. 1(a) shows the UMAP of the SVD reduced data without any transformation as the baseline. Cell groups from different major cell types were separated well based on the true labels, as seen Fig. 1(b). Also as shown in Figs. 1(a-c), excitatory neuron cells as well as other cell types have formed distinct clusters. This emphasizes the impact of the inherent sparsity of scATAC-seq on cell states. All the scenarios are evaluated and illustrated in Figs. 1(d,h). The best transformation method based on ARI and NMI was found to be $TF \times log(IDF)$. All TF-IDF methods showed better or similar performance compared to the baseline, except for the entropy-based transformation (Method 5). In addition, as seen from Figs. 1(d,h) removing the first LSI component increased ARI. This is consistent with the previous finding that the first LSI component in single-cell ATAC-seq data is correlated with sequencing depth. Thus, the first LSI component was removed from the downstream analysis [5,7].

*B. Top-weighted regions from SVD factorization reveal cell-specific pathways and transcription factors*

By extracting genomic regions that have high loci loadings on a given LSI component (D section in Methods), the top regions were found, as illustrated in Figs. 2(a,c) for two different LSI components. Some of the LSI components also showed cell-specific cell loadings. For instance, based on the dataset in this study, the 3rd and 5th LSI components showed cell-specific loadings, as seen in Figs. 2(b,d). The majority of cells with high loading in the 3rd LSI are microglial cells (MG), as shown in Figs. 1(f) and 3(b). This raises the question if the top features (loci) for the 3rd LSI component are glial-specific. Fig. 2(e) shows the top 8 enriched motifs and the TFs associated with these loci including transcription factor bind motifs corresponding to MEF2D and MEF2B. Myocyte Enhancer Factor-2 (MEF2) is the family of transcription factors known to display enrichment for motifs specific to microglial cells [17]. The enriched terms based on the Molecular Signatures Database (MSigDb) pathway show enriched terms ($p_{val} < 10^{-4}$) for transcription of clock-related genes REV-ERBa and RORA in the circadian system [18]. Top cells in the 5th LSI component are dominantly astrocyte- and microglial-specific, *see* Fig. 2(f). Based on the motif enrichment analysis, most of the enriched motifs are associated with HOX and FOX transcription factor families, as shown in Fig. 2(g). FOX plays an important role in regulating astrocytes in the developing brain, and HOX is a homeodomain specific to organ development. It is well known that astrocytes and microglia cells are vital in synapse formation and remodeling [19], which supports our finding from the enriched terms ($p_{val} < 10^{-6}$) for GO cellular components, including synapse and post-synapse terms.

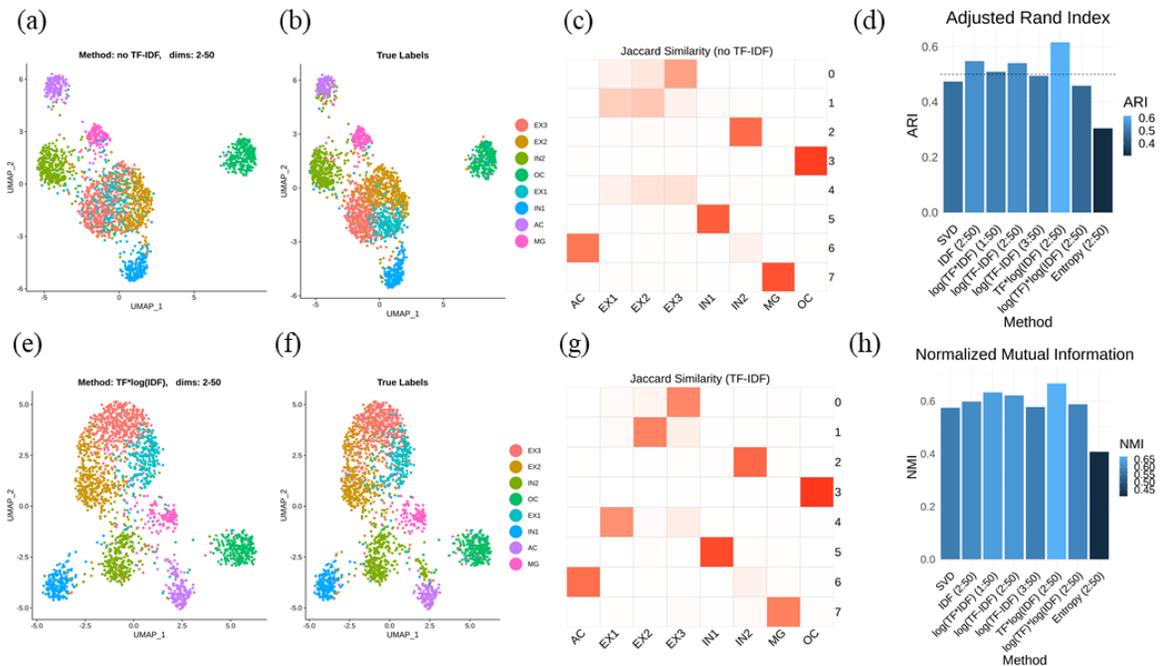

Fig. 1. Clustering evaluation on the mouse forebrain data based on different transformation methods: Without any TF-IDF transformation, the UMAP of the reduced data based on (a) clustering and (b) true cell labels are illustrated which serves as a baseline. For the selected $TF \times log(IDF)$ transformation, (e) clustered and (f) true labels are shown in UMAPs. Heatmap of Jaccard similarity for cells between true labels (rows) and clusters (columns) are depicted in (c) for non-transformed data and (d) TF-IDF transformed data after SVD reduction. (d) shows the adjusted rand index and (h) the normalized mutual information for all the methods scenarios.

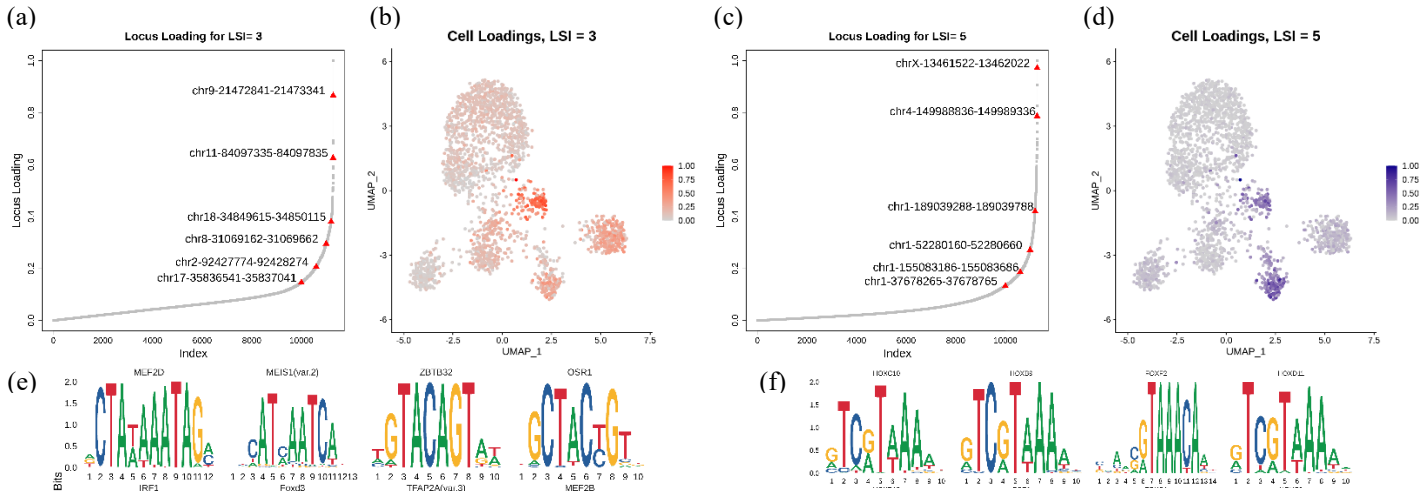

Fig. 2. Feature analysis of the mouse forebrain dataset based on SVD transformation: (a) and (c) show scaled feature loadings of the extracted important features, (b) and (d) show cell loadings on the projected data (UMAP), (e) and (f) show sequence logos (in bits) of the enriched motifs of the extracted regions from the LSI components 3 and 5, respectively.

### C. Matrix factorization models benefit from TF-IDF transformation methods

Using the best-performing TF-IDF method (Method 4, Table I) and combined with the lsNMF and NMF methods, we observed significant improvement in cell state detection. Further, the NMF method without the transformation labeled the majority of cells as excitatory neuron cells 3. However, by applying the transformation, most of the clusters, except for a few excitatory neuron cells, were correctly clustered. lsNMF method without transformation shows most of the cells labeled as inhibitory cells 2, and after applying the transformation, the performance significantly increases (Table II). Moreover, the two matrix factorization models performed better than the SVD-based method only. However, one major drawback here is the EX1 and EX2 cell groups, as they are missing in all these methods. EX1 and EX2 are hard to cluster based on a linear dimension reduction technique. So, a nonlinear dimension reduction method may help a more reliable clustering and downstream analysis.

### D. Autoencoders benefit from both TF-IDF transformation methods and nonlinearity for dimension reduction

Table II summarizes the performance of the variational autoencoders: general autoencoder, sparse autoencoder, and variational autoencoder, with and without transformation. The general autoencoder worked better than matrix factorization models that have no transformations. However, it still suffers from very low clustering quality for EX1 and EX2 cells, although the transformation has increased all the performance metrics compared to no transformation. The sparse autoencoder has failed to predict astrocyte cells compared to the general autoencoder. Hence, a more sophisticated autoencoder may resolve this issue. Compared to other autoencoders or factorization methods, VAE without TF-IDF has outperformed the rest (Table II). Also, more EX1 and EX2 cells are predicted correctly. However, EX3 cells are surprisingly not correctly predicted (Fig. 3(a)). TF-IDF transformation overcame the issue of excitatory neuron mislabeling, as in Fig. 3(b), most cell states are correctly predicted, and the combination of TF-IDF transformation and VAE outperformed all other studied methods (Table II). The last class of autoencoder analyzed was the stacked autoencoder. For the mouse dataset, which is not too large, the algorithm either terminated prematurely with not enough iterations or never finished even after 100 times more of the time spent for any other autoencoders. Since this structure is neither feasible for our dataset nor scalable to any larger dataset, we decided to disregard the analysis for this case.

### E. TF-IDF methods are scalable

We applied the same comparison to the second dataset, which is a much larger dataset (PBMC, 10k cells and 131k loci, see Fig. 4(a)) and investigated the effect of the transformation. The TF-IDF Methods (2)-(5) combined with SVD successfully improved NMI and ARI in a reasonable time using SLM clustering (Fig. 4 (b,c)). To investigate autoencoder performance, we applied the entropy-based transformation as it was the best method in retrieving cell states, and then clustered cells using kNN. Two separate datasets were prepared to investigate the effect of the transformation on detecting rare cell types: (1) major cell types (excluding rare cell types) and (2) all cell types. Since the training of the large data was computationally expensive, the SVD reduced data instead was fed to the autoencoders to increase efficacy. From Table III, improved clustering performance when using TF-IDF is observed for both case studies.

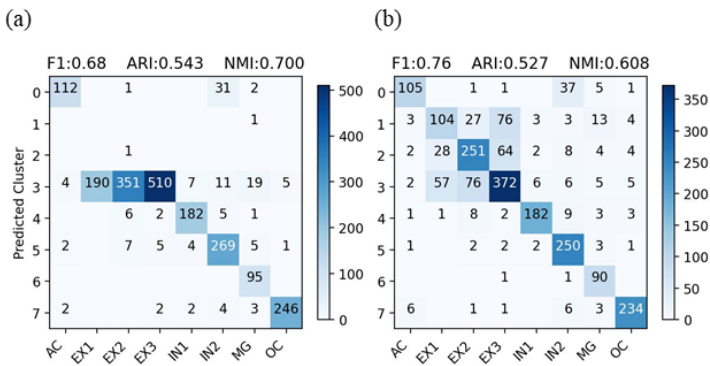

Fig. 3. Clustering evaluation on the variational autoencoder on the mouse forebrain dataset. (a) shows evaluation based on non-transformed clustered cells for (a) and (b) TF-IDF transformed data.

Table II. Performance metrics for all methods

|  | ARI | | NMI | | F1 Score | |
|---|---|---|---|---|---|---|
| TF-IDF? | No | Yes | No | Yes | No | Yes |
| SVD | 0.07 | 0.12 | 0.18 | 0.35 | 0.29 | 0.40 |
| NMF | 0.08 | 0.47 | 0.25 | **0.68** | 0.36 | 0.68 |
| lsnmf | 0.11 | 0.45 | 0.29 | 0.67 | 0.39 | 0.67 |
| AE | 0.17 | 0.37 | 0.33 | 0.51 | 0.40 | 0.54 |
| spAE | 0.18 | 0.28 | 0.33 | 0.42 | 0.39 | 0.47 |
| VAE | **0.54** | 0.53 | **0.70** | 0.61 | **0.68** | 0.76 |

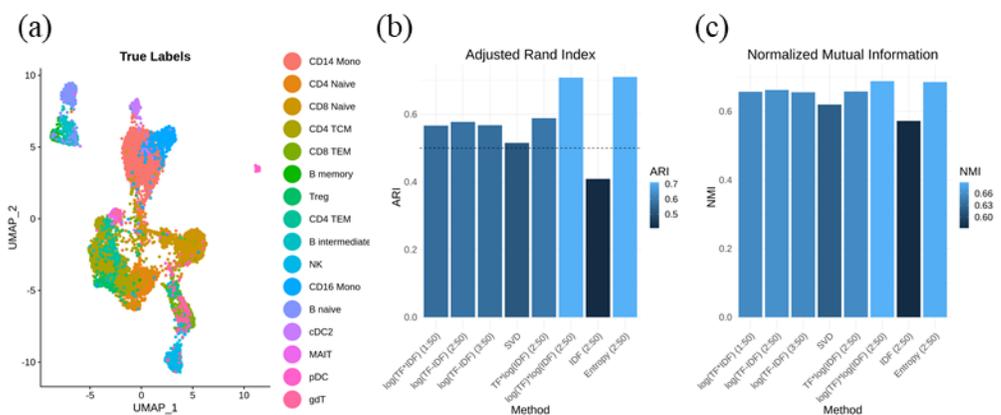

Fig. 4. TF-IDF performance evaluation on the human PBMC data: (a) UMAPs show cell labels of the gold standard annotations. Cell labels are evaluated for (b) adjusted rand index and (c) normalized mutual information.

## IV. CONCLUSIONS AND FUTURE DIRECTIONS

TF-IDF methods show great potential for scATAC-seq data analysis. Autoencoders combined with TF-IDF transformation, especially VAEs, show stronger discrimination among subpopulations of cells. It appears that linear models, such as NMF benefit more from the TF-IDF transformation compared to nonlinear dimension reduction techniques. Autoencoders benefit less from TF-IDF compared to factorization methods. However, retrieving informative, original features (peaks) is harder. One future direction to explore is the feature analysis from autoencoders. Determining the best TF-IDF transformation through comprehensive evaluation using various large-scale data is another direction, as there has been no agreement on a "gold transformation method". The choice of clustering also matters; the SLM modularity optimization on kNN graph is known to be more effective than simple k-means clustering. Identification of rare cell types is still challenging and needs further attention.

## REFERENCES

[1] J. D. Buenrostro et al., "Single-cell chromatin accessibility reveals principles of regulatory variation," *Nature,* vol. 523, no. 7561, pp. 486-90, Jul 23 2015, doi: 10.1038/nature14590.
[2] C. Bravo Gonzalez-Blas et al., "cisTopic: cis-regulatory topic modeling on single-cell ATAC-seq data," *Nat Methods,* vol. 16, no. 5, pp. 397-400, May 2019, doi: 10.1038/s41592-019-0367-1.
[3] T. Stuart et al., "Comprehensive Integration of Single-Cell Data," *Cell,* vol. 177, no. 7, pp. 1888-1902 e21, Jun 13 2019, doi: 10.1016/j.cell.2019.05.031.
[4] C. Wang et al., "Integrative analyses of single-cell transcriptome and regulome using MAESTRO," *Genome Biol,* vol. 21, no. 1, p. 198, Aug 7 2020, doi: 10.1186/s13059-020-02116-x.
[5] T. Stuart, A. Srivastava, S. Madad, C. A. Lareau, and R. Satija, "Single-cell chromatin state analysis with Signac," *Nat Methods,* vol. 18, no. 11, pp. 1333-1341, Nov 2021, doi: 10.1038/s41592-021-01282-5.
[6] "Cell Ranger ATAC Algorithms Overview." 10X Genomics. https://support.10xgenomics.com/single-cell-atac/software/pipelines/latest/algorithms/overview#clustering (accessed 8/22/2022, 2022).
[7] D. A. Cusanovich et al., "A Single-Cell Atlas of In Vivo Mammalian Chromatin Accessibility," *Cell,* vol. 174, no. 5, pp. 1309-1324 e18, Aug 23 2018, doi: 10.1016/j.cell.2018.06.052.
[8] G. Salton and C. T. Yu, "On the construction of effective vocabularies for information retrieval," *SIGPLAN Not.,* vol. 10, no. 1, pp. 48–60, 1973, doi: 10.1145/951787.951766.
[9] J. Baglama, "IRLBA: Fast Partial Singular Value Decomposition Method," in *Handbook of Big Data*, 2016.
[10] Y. Huang, Y. Li, Y. Liu, R. Jing, and M. Li, "A Multiple Comprehensive Analysis of scATAC-seq Based on Auto-Encoder and Matrix Decomposition," *Symmetry,* vol. 13, no. 8, p. 1467, 2021.
[11] S. Preissl et al., "Single-nucleus analysis of accessible chromatin in developing mouse forebrain reveals cell-type-specific transcriptional regulation," (in eng), *Nat Neurosci,* vol. 21, no. 3, pp. 432-439, Mar 2018, doi: 10.1038/s41593-018-0079-3.
[12] A. J. Hill. "Dimensionality Reduction for scATAC Data." http://andrewjohnhill.com/blog/2019/05/06/dimensionality-reduction-for-scatac-data/ (accessed 23/8/2022).
[13] M. W. Berry and M. Browne, *Understanding Search Engines: Mathematical Modeling and Text Retrieval (Software, Environments, Tools), Second Edition.* Society for Industrial and Applied Mathematics, 2005.
[14] C. Y. McLean et al., "GREAT improves functional interpretation of cis-regulatory regions," *Nat Biotechnol,* vol. 28, no. 5, pp. 495-501, May 2010, doi: 10.1038/nbt.1630.
[15] L. Hubert and P. Arabie, "Comparing partitions," *Journal of Classification,* vol. 2, no. 1, pp. 193-218, 1985/12/01 1985, doi: 10.1007/BF01908075.
[16] A. Strehl and J. Ghosh, "Cluster Ensembles --- A Knowledge Reuse Framework for Combining Multiple Partitions," *J. Mach. Learn. Res.,* vol. 3, pp. 583-617, 2002.
[17] A. Machado Xavier, S. Belhocine, and D. Gosselin, "Essential contributions of enhancer genomic regulatory elements to microglial cell identity and functions," *WIREs Systems Biology and Medicine,* vol. 11, no. 5, p. e1449, 2019, doi: 10.1002/wsbm.1449.
[18] J. Lee et al., "Inhibition of REV-ERBs stimulates microglial amyloid-beta clearance and reduces amyloid plaque deposition in the 5XFAD mouse model of Alzheimer's disease," *Aging Cell,* vol. 19, no. 2, p. e13078, Feb 2020, doi: 10.1111/acel.13078.
[19] I. D. Vainchtein and A. V. Molofsky, "Astrocytes and Microglia: In Sickness and in Health," *Trends Neurosci,* vol. 43, no. 3, pp. 144-154, Mar 2020, doi: 10.1016/j.tins.2020.01.003.